\documentclass[twocolumn,showpacs,aps,prl,superscriptaddress]{revtex4}

\newcommand{\hz}{\ensuremath{h^0}\xspace}
\newcommand{\DE}{\ensuremath{\Delta E}\xspace}
\newcommand{\mES}{\ensuremath{m_\mathrm{ES}}\xspace}
\def\calC{{\ensuremath{\cal C}}\xspace}
\newcommand{\dm}{\ensuremath{\Delta m}\xspace}
\newcommand{\dt}{\ensuremath{\Delta t}\xspace}
\newcommand{\Btag}{\ensuremath{B_\mathrm{tag}}\xspace}

\newcommand{\sigmadt}{\ensuremath{\sigma_{\Delta t}}\xspace}
\def\etal{{\it et al.}}

\usepackage{graphicx}
\usepackage{dcolumn}
\usepackage{amsmath}
\usepackage{epsfig}
\usepackage{bigstrut}

\input babarsym.tex

\newcommand{\BABARPubYear}    {07}
\newcommand{\BABARPubNumber}  {006}
\newcommand{\SLACPubNumber} {12380}

\def\figurebox#1#2#3{%
    \def\arg{#3}%
    \ifx\arg\empty
    {\hfill\vbox{\hsize#2\hrule\hbox to #2{\vrule\hfill\vbox to #1{\hsize#2\vfill}\vrule}\hrule}\hfill}%
    \else
    {\hfill\epsfbox{#3}\hfill}%
    \fi}

\begin{document}

\preprint{\babar-PUB-\BABARPubYear/\BABARPubNumber}
\preprint{SLAC-PUB-\SLACPubNumber}

\begin{flushleft}
\begin{minipage}{\textwidth}
\babar-PUB-\BABARPubYear/\BABARPubNumber\\
SLAC-PUB-\SLACPubNumber
\end{minipage}
\end{flushleft}

\title{
{\large \bf
Measurement of the Time-Dependent {\boldmath \CP} Asymmetry 
in {\boldmath $\Bz\ra D^{(*)}_{\CP}\hz$} Decays}
}

%
\author{B.~Aubert}
\author{M.~Bona}
\author{D.~Boutigny}
\author{Y.~Karyotakis}
\author{J.~P.~Lees}
\author{V.~Poireau}
\author{X.~Prudent}
\author{V.~Tisserand}
\author{A.~Zghiche}
\affiliation{Laboratoire de Physique des Particules, IN2P3/CNRS et Universit\'e de Savoie, F-74941 Annecy-Le-Vieux, France }
\author{J.~Garra~Tico}
\author{E.~Grauges}
\affiliation{Universitat de Barcelona, Facultat de Fisica, Departament ECM, E-08028 Barcelona, Spain }
% ...
\collaboration{The \babar\ Collaboration}
\noaffiliation

\date{\today}

\begin{abstract}
We report a measurement of the time-dependent \CP-asymmetry
parameters \calS and \calC in color-suppressed $\Bz \ra D^{(*)0}\hz$
decays, where \hz
is a \piz, $\eta$, or $\omega$ meson, and the \Dz
decays to one of the \CP eigenstates $\Kp\Km$, $\KS\piz$, or $\KS\omega$. The
data sample consists of $383\times 10^{6}$ $\Upsilon(4S)\ra\BB$ decays
collected with the \babar\ detector at the PEP-II asymmetric-energy \B factory
at SLAC. The results are $\calS=
-0.56 \pm 0.23 \pm 0.05$ and $\calC= -0.23\pm 0.16\pm 0.04$, where the
first error is statistical and the second is systematic.
\end{abstract}


\maketitle

Measurements of time-dependent \CP asymmetries in \Bz meson decays,
through the interference between decays with and without \Bz-\Bzb mixing,
have provided stringent tests on the mechanism of \CP violation in the
standard model (SM).
The time-dependent \CP asymmetry amplitude $\sin2\beta$ has been
measured with high precision in the $\b\ra\ccbar \s $ decay
modes~\cite{Sin2B_ccK}, where $\beta=
-\mathrm{arg}(V_{cd}V_{cb}^*/V_{td}V_{tb}^*)$ is a phase in the
Cabibbo-Kobayashi-Maskawa (CKM) quark-mixing matrix~\cite{CKM}.

In this Letter, we present a measurement of the time-dependent \CP
asymmetry in
\Bz meson decays to a neutral $D$ meson and a light neutral meson
through a $\b\ra\c\ubar \d$ color-suppressed tree amplitude. 
Interference between decay amplitudes
with and without \Bz-\Bzb mixing contribution occurs if
the neutral $D$ meson decays to a \CP eigenstate.
The measured time-dependent asymmetry is expected to be different from
$\sin2\beta$ measured in the charmonium modes due to the subleading
amplitude $\b\ra\u\cbar \d$, which has a different weak phase. This amplitude
is suppressed by $V_{ub}V^*_{cd}/V_{cb}V^*_{ud} \simeq 0.02$
relative to the leading diagram.
Therefore the deviation is expected to
be small in the SM~\cite{Grossman:1996ke,Fleischer}.

Many other decay modes 
that have significant contribution from loop diagrams
have been studied~\cite{cpviolation} to constrain or discover new physics
due to unobserved heavy 
particles in the loop diagrams in \B decays.
This kind of new physics would not affect the decays presented in this
Letter because only tree diagrams contribute to these modes.
However, $R$-parity-violating ($\not\!\!\!R_p$)
supersymmetric processes~\cite{Grossman:1996ke,RVprocess}
 could enter at tree level in
these decays, leading to a deviation from the SM prediction.

The analysis uses a data sample of $348~$\invfb, which corresponds to
$(383\pm 4)\times 10^{6}$ \FourS decays into
$\BB$ pairs collected with the \babar\
detector at the asymmetric-energy $\epem$ PEP-II collider.
The \babar\
detector is described in detail elsewhere~\cite{babar}.
We use the {\ttfamily GEANT4} simulation toolkit~\cite{geant} to
simulate interactions of particles traversing the \babar\ detector,
and to take into account the varying detector conditions and beam
backgrounds.

We fully reconstruct \Bz mesons~\cite{conjugate} decaying into a \CP
eigenstate in the following channels:
$D^{(*)0} \piz$ ($\Dz \to K^+ K^-,~\KS \omega$)~\cite{d0tocp} and
$D^{(*)0} \eta$ ($\Dz \to K^+ K^-$)  with
$\Dstarz\to\Dz\piz$, and
$\Dz \omega$ ($\Dz \to K^+ K^-,~\KS \omega,~\KS \piz$).
From the remaining particles in
the event, the vertex of the other \B meson, \Btag, is reconstructed and
its flavor is identified (tagged). The proper decay time difference $\dt=
t_{\CP}- t_\mathrm{tag}$, between the signal \B ($t_{\CP}$) and \Btag
($t_\mathrm{tag}$) is determined from the measured distance between the
two \B decay vertices projected onto the boost axis
and the boost ($\beta\gamma= 0.56$) of the
center-of-mass (c.m.) system. The \dt distribution is
given by:
\begin{align}
F_\pm(\dt) & = \frac{e^{-|\dt|/\tau}}{4\tau} [ 1\mp \Delta w \pm
 \label{eq:dt}\\
 & (1-2w) (\eta_f \calS \sin(\dm\dt) - \calC \cos(\dm\dt)) ]\,,  \notag
\end{align}
where the upper
(lower) sign is for events with \Btag being identified as a \Bz (\Bzb),
 $\eta_f$ is the \CP eigenvalue of the final state,
\dm is the \Bz-\Bzb mixing frequency, $\tau$ is the mean lifetime of
the neutral \B meson, the mistag parameter $w$ is the probability of
incorrectly identifying the flavor of \Btag, and $\Delta w$ is the difference
of $w$ for \Bz and \Bzb.
The neural-network based tagging algorithm~\cite{tag}
has six mutually exclusive categories and a measured total effective
tagging efficiency of $(30.4\pm 0.3)\%$.
Neglecting CKM-suppressed decay amplitudes, we expect the \CP violating
parameters
$\calS = -\sin 2\beta $ and $\calC = 0$ in the SM.

The event selection criteria are determined by maximizing the expected signal
significance based on the
simulation of signal and generic decays of \BB and $\epem \ra
\qqbar$ ($q = u, d, s, c$) continuum events.
The selection requirements vary by mode due to different 
signal yields and background levels.

A pair of energy clusters in the electromagnetic calorimeter (EMC),
isolated 
from any charged tracks and with a lateral shower shape consistent with
photons, is considered as a $\piz$ candidate if both cluster energy deposits
exceed 30\mev
and the invariant mass of the pair is between 100 and 160\mevcc.
Charged tracks are considered as pions, except for those used in
$\Dz\ra\Kp\Km$ reconstruction, where the kaons must be consistent with the
kaon hypothesis~\cite{s2bprd}.
We reconstruct $\eta$ mesons in
$\gamma\gamma$ and $\pip\pim\piz$ modes.
Each photon is required to have an energy
exceeding 100\mev and, when combined with any other photon in the event, 
to not have an invariant mass
within 5\mevcc of the \piz nominal mass~\cite{nominal}.
The invariant mass is required to be within approximately 30\mevcc (8\mevcc)
of the $\eta$
nominal mass for $\eta\to\gamma\gamma$ ($\eta\to\pip\pim\piz$).
Both \piz and $\eta\to\gamma\gamma$ candidates are kinematically fitted with
their invariant masses constrained at their respective nominal values.
The $\omega\ra\pip\pim\piz$ candidates are
accepted if the invariant mass is within
approximately 22\mevcc of the nominal $\omega$ mass, depending on the \Dz
decay mode. 
The $\KS\ra\pip\pim$ candidates are required to have an invariant mass within
$10$\mevcc of the $\KS$ nominal mass and $\chi^2$ probability of forming a
common vertex greater than 0.1\%.
The distance between the \KS decay vertex and the primary interaction point
projected on the plane perpendicular to the beam axis is required to be
greater than twice its measurement uncertainty.

The vector meson $\omega$ is fully polarized in $\Dz\to \KS \omega$
decays. Two angular distributions of the $\omega$ decay are used 
to discriminate against background: 
(a) $\cos \theta_N^D$, defined in the $\omega$ rest frame, the cosine of the
angle between the \Dz direction and the normal to the decay plane of
$\omega\ra\pip\pim\piz$, and 
(b) $\cos \theta_D^D$, the cosine of the angle between the direction of one
pion in the rest frame of the remaining pion pair and the direction of the
pion pair. The signal are distributed according to
$\cos^2\theta_N^D$ and $1 - \cos^2\theta_D^D$, while the background
distributions are nearly uniform. We require
$|\cos \theta_N^D|>0.4$ and $|\cos \theta_D^D|<0.9$.

For the $\Dz$ in $\Dstarz\to\Dz\piz$, the invariant mass of the \Dz candidate
is required to be within 30\mevcc of the world-average \Dz mass. 
For the $\Dz$ in $\Bz\ra\Dz\hz$, the invariant mass window is tightened,
ranging from $\pm14$ to $\pm29$\mevcc, depending on the mode.
In both cases the \Dz is
kinematically fitted with its mass constrained at its nominal value.
 The invariant mass difference between $\Dstarz$ and \Dz candidates
is required to be within $\pm 2.7$\mevcc\ of the nominal
value. For $\Bz \to \Dstarz \piz$ with $\Dz \to \KS \omega$, we require 
$|\cos\theta_H^*|>0.4$, where $\theta_H^*$ is the angle between 
the momenta of the $\Bz$ and the $\piz$ from the $\Dstarz$ in the $\Dstarz$
rest frame.

The signal is characterized by the kinematic variables
$\mES= \sqrt{(s/2 + {\mathbf p}_0 \cdot {\mathbf p}_B)^2 / E_0^2 -
{\mathbf p}_B^2}$, and $\DE = E^*_{\B} - E_{\rm beam}^*$,
where the asterisk denotes the values evaluated in the c.m. frame, the
subscripts $0$, beam and \B denote the \epem system, the beam and the \B
candidate, respectively, and $\sqrt{s}$ is the c.m.~energy.
We require $\mES>5.23\gevcc$.
The \DE distribution for signal events is asymmetric and varies by decay
mode. 
Depending on the mode, the lower (upper) boundary of the \DE selection window
varies 
from $-95$ to $-35$\mev ($+35$ to $+85$\mev).
The reconstructed $|\dt|$ and its uncertainty \sigmadt are required
to satisfy $|\dt|<15$~ps and $\sigmadt<2.5$~ps.

The background from continuum $\qqbar$ production  is
suppressed based on the event topology. 
In the c.m.~frame,
the \B mesons are produced nearly at rest and decay isotropically, while the
quarks in the process $\epem\to\qqbar$ are
produced with large relative momentum and result in a jetlike topology.
The ratio of the second to
zeroth order Fox-Wolfram moments~\cite{foxwolfram}, 
determined from all charged tracks and clusters in the EMC with energy greater
than 30\mev, must be less than 0.5. The $\qqbar$
background is further suppressed by a Fisher discriminant $\cal
F$~\cite{fisher}, constructed with the following variables,
evaluated in the c.m.~frame:
(a) $L_{2}/L_{0}$ where $L_{i}= \sum_j p^*_j
|\cos\theta^*_j|^i$, summed over the remaining particles in the
event after removing the daughter particles from the \Bz,
$p^*_j$ is the momentum of particle $j$ and
$\theta^*_j$ is the angle of the momentum with respect to the \Bz
thrust axis~\cite{thrust};
(b) $|\cos\theta^*_T|$, where $\theta^*_T$ is the angle between the \Bz thrust
axis and the thrust axis of the rest of the event;
(c) $|\cos^2\theta^*_{\B}|$, where $\theta^*_{\B}$ is the angle between the
beam direction and the direction of the \Bz;
(d) total event thrust magnitude; and
(e) total event sphericity~\cite{sphericity}.

For $\Bz\to \Dz\omega$ decays, we add two angular variables to $\cal F$:
$\cos \theta_N^B$ and $\cos \theta_D^B$, analogous to $\cos\theta_N^D$
and $\cos \theta_D^D$ in $\Dz\to \KS \omega$.
The signal distributions for the \Bz system 
are the same as those in the \Dz system. The background distributions
are close to $2 - \cos^2 \theta_N^B$ and uniform in $\cos\theta_D^B$.
The requirement on $\cal F$ depends on the background level in each mode;
the signal selection (background rejection) efficiency is 60\%--86\% 
(72\%--94\%).

Within each reconstructed decay chain, the fraction of events that have more
than one candidate ranges from less than 1\% to about 10\%, depending on the
mode. We select one candidate with the most signal-like
Fisher discriminant value for each mode. 
A total of 1128 events are selected, of which 751 are tagged 
(the absolute value of the flavor-tagging neural-network output greater
than 10\% of the maximum).

\begin{figure}[tb]
\begin{center}
\includegraphics[width=0.235\textwidth]{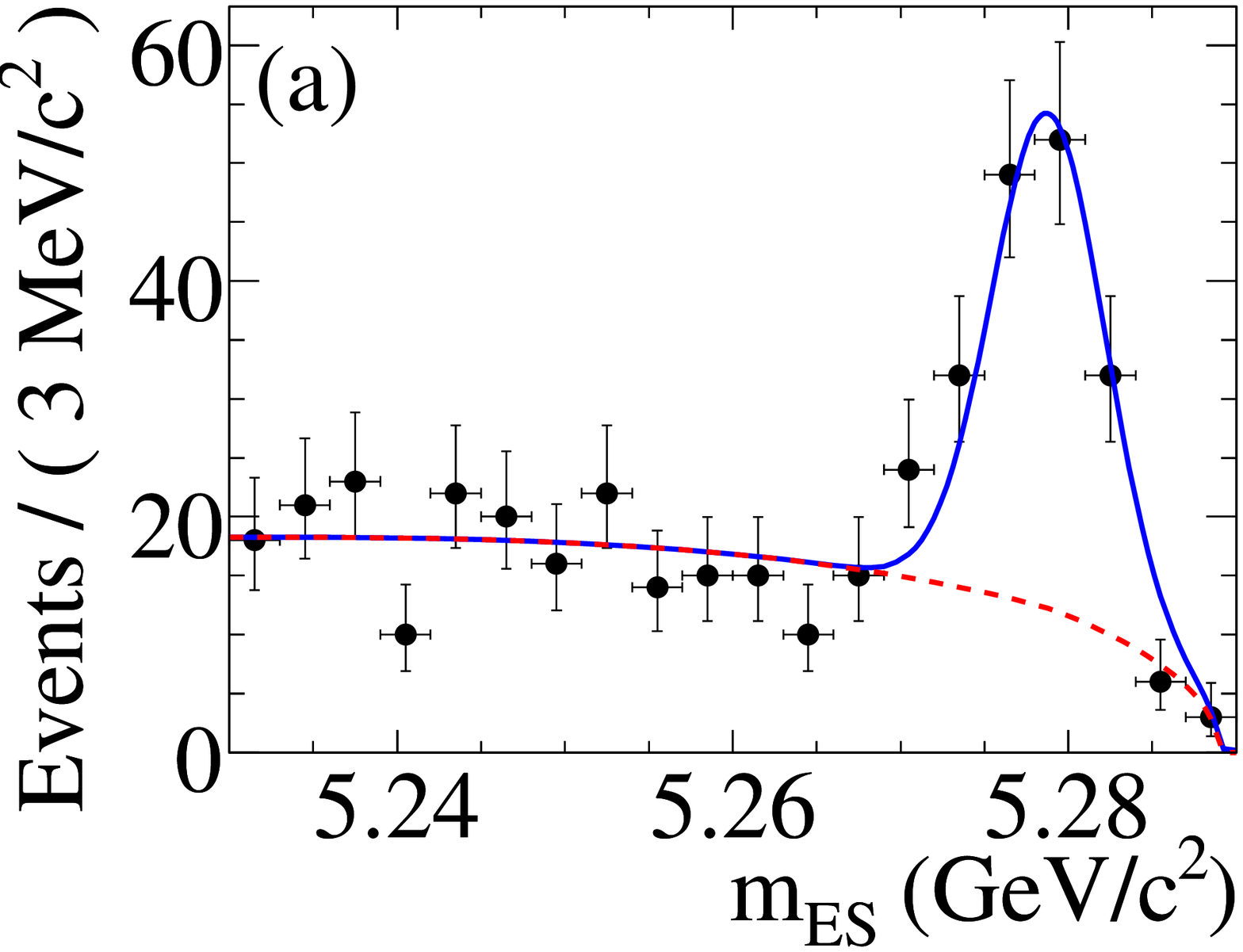}
\includegraphics[width=0.235\textwidth]{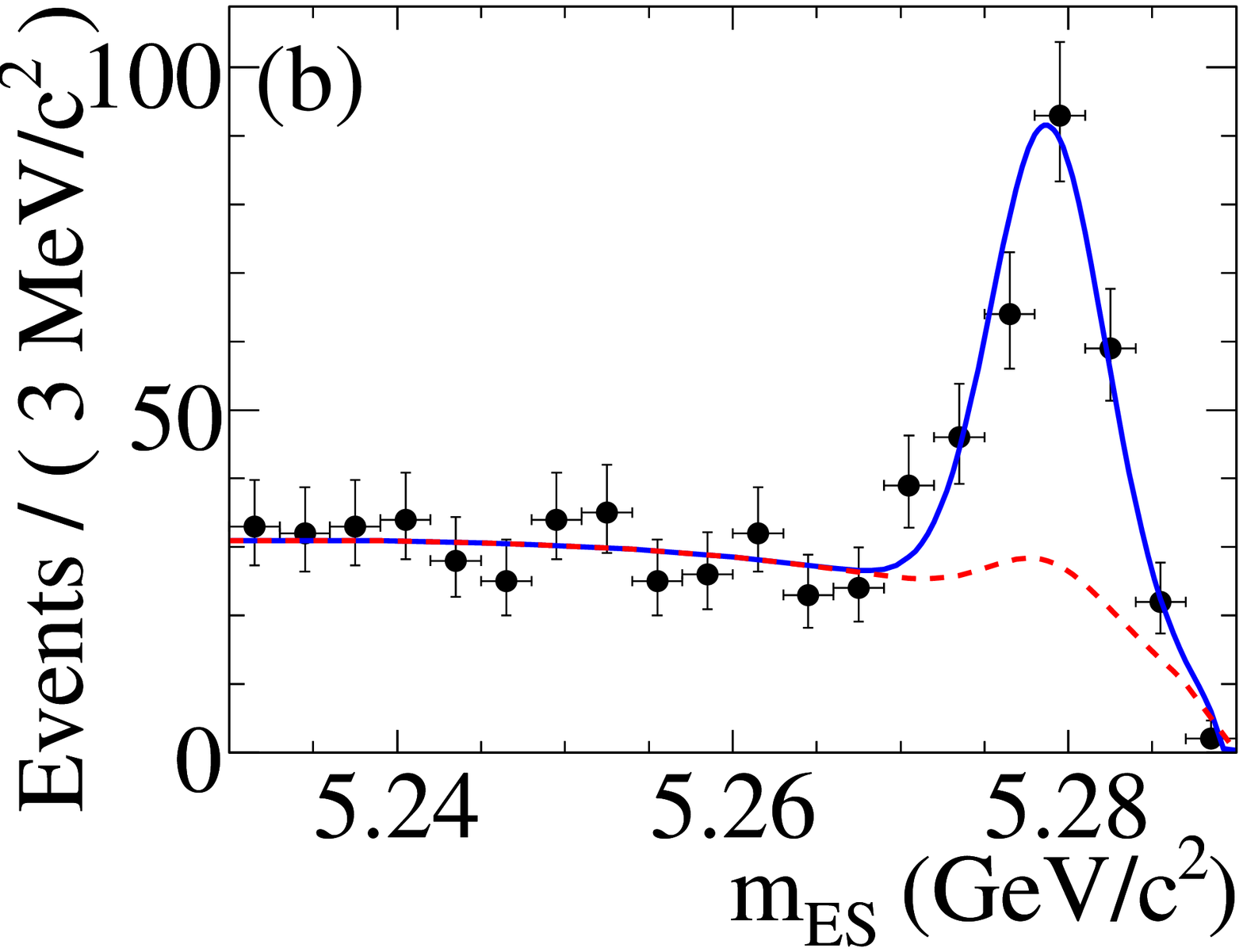}
\end{center}
\caption{
The \mES distributions with a fit to (a) the \CP-even and (b) the \CP-odd modes combined in the data.
The solid curve represents the overall PDF projection and the dashed curve represents the background.
}
\label{fig:datafitmes-even-odd}
\end{figure}

The signal and background yields are determined by a fit to the \mES 
distribution using a Gaussian distribution for the signal peak and
a threshold function~\cite{argus} for the combinatorial background.
We obtain $340\pm 32$ signal events ($259\pm 27$ tagged).
The contribution from each mode 
is shown in Table~\ref{tab:fitmes}, and the \mES
distributions are shown in Fig.~\ref{fig:datafitmes-even-odd}.
We investigate potential backgrounds that might peak in the 
\mES signal region by studying data in the \Dz mass sideband 
(outside a window of $\pm 3$ standard deviations of the mass peak)
and simulated $\epem\to\BB$ events.
We estimate that $(0.8\pm 2.6)$\% of the \CP-even signal yield
and $(5.4\pm 2.2)$\% of the \CP-odd signal yield are background, based
on the 
simulation. Approximately half of the peaking background found in simulation is from
$\Bm\ra\Dz\rho^-(\ra \piz\pim)$ with a low momentum \pim. Other sources
include $\Bz\ra \pip\pim\piz$ and $\Bz\ra D^{(*)0}\hz$, with \Dz
decaying to a flavor eigenstate, e.g., $\Km\pip$.
We find that the peaking background from the \Dz mass sideband data in 
\CP-even modes is consistent with the simulation.
For \CP-odd modes, we find a larger
peaking component in \Dz sideband data than expected from
simulation. 
Therefore we increase the estimated total peaking background fraction for
\CP-odd events to $(11\pm 6)\%$ to account for the excess found in the \Dz
sideband data. We estimate that $65\%$ of the peaking background arises from
charmless decays with potentially large \CP-violating
asymmetries.  We account for this possibility in the 
systematic uncertainty.

\begin{table}[tb]
\caption{Signal yields. Uncertainties are  statistical only. The \CP parity of
  the \Dz is indicated in the column of $D_{\CP}$. The combined value is from
  a simultaneous fit to all modes.
}
\begin{center}
\begin{tabular*}{0.475\textwidth}{@{\extracolsep{\fill}}l@{\extracolsep{\fill}}c@{\extracolsep{\fill}}c|@{\extracolsep{\fill}}l@{\extracolsep{\fill}}c@{\extracolsep{\fill}}c}
\hline\hline
\multicolumn{3}{c|}{$\eta_f=+1$ (\CP even)} &\multicolumn{3}{c}{$\eta_f=-1$ (\CP odd)} \\
\hline
 Mode & $D_{\CP}$ & $N_\mathrm{signal}$ &
 Mode & $D_{\CP}$ & $N_\mathrm{signal}$\\
\hline
$\Dz_{\KS\omega}\piz$ &  $-$ &  $ 26.2 \pm 6.3$ &
$\Dz_{KK}\piz$ & $+$ & $ 104 \pm 17 $ \\
$\Dz_{\KS\piz}\omega$ &  $-$ & $ 40.0 \pm 8.0$ &
$\Dz_{KK}\eta_{\gamma\gamma}$ & $+$  & $ 28.9 \pm 6.5$ \\
$\Dz_{\KS\omega}\omega$ & $-$ & $ 23.2 \pm 6.8$ &
$\Dz_{KK}\eta_{3\pi}$ & $+$ & $ 14.2 \pm 4.7$ \\
$\Dstarz_{KK}\piz$ & $+$ & $ 23.2 \pm 6.3$ &
$\Dz_{KK}\omega$ & $+$ & $ 51.2 \pm 8.5$ \\
$\Dstarz_{KK}\eta_{\gamma\gamma}$ & $+$ & $ 9.8 \pm 3.5$ &
$\Dstarz_{\KS\omega}\piz$ & $-$ & $ 5.5 \pm 3.3$ \\
$\Dstarz_{KK}\eta_{3\pi}$ & $+$ & $6.8 \pm 2.9$ \\
\hline
Combined &  & \multicolumn{1}{c}{$131 \pm 16$\;\;} & & & $ 209 \pm 23$ \\
\hline
Total & \multicolumn{4}{c}{} & $ 340 \pm 32$ \\
\hline\hline
\end{tabular*}
\end{center}
\label{tab:fitmes}
\end{table}

In order to extract \CP violating parameters $\calS$ and $\calC$, we fit the
\mES and \deltat distributions of the 751 tagged events
using a two-dimensional probability density
function (PDF) that contains three components:
signal, peaking background and combinatorial background. 
The \mES distribution is described in the previous paragraph. 
Its parameters are free in the fit.
The
peaking background is assumed to have the same \mES shape as the signal.
The signal decay-rate distribution shown in Eq.~(\ref{eq:dt})
accounts for dilution due to an incorrect assignment of the flavor of
\Btag, and is convolved with a sum of three Gaussian
distributions, parameterizing the core, tail and outlier parts of the \dt
resolution function~\cite{s2bprd}. 
The widths and biases of the core and tail
Gaussians are scaled by \sigmadt. 
The biases are nonzero to account for the charm meson flight from the \Btag
vertex. 
The outlier Gaussian has a fixed mean
(0~ps) and width (8~ps) to account for poorly-reconstructed
decay vertices. The mistag parameters and the
resolution function are determined from a large data control sample of
$\Bz\ra D^{(*)-}h^{+}$ decays, where $h^{+}$ is a $\pip$, $\rho^+$, or
$a_1^{+}$ meson. The \Bz lifetime and mixing frequency are taken
from~\cite{pdg}. 

We use an exponential decay to model the \deltat PDF of the peaking
background. We account for possible \CP asymmetries in the systematic
uncertainty. 
The \deltat PDF for combinatorial background consists of a term
with zero lifetime to account for the \qqbar contribution, and an oscillatory
term whose effective lifetime and 
oscillatory coefficients are free parameters in the fit to account
for possible \CP asymmetry in the background.
The sum of a core Gaussian and an
outlier Gaussian is sufficient to model the resolution function.
The combinatorial
background parameters are determined predominately by the events in
the \mES sideband. The final PDF has 25 free parameters for fitting to all
modes and tagging categories simultaneously.

\begin{figure}[tb]
\begin{center}
\includegraphics[width=0.235\textwidth]{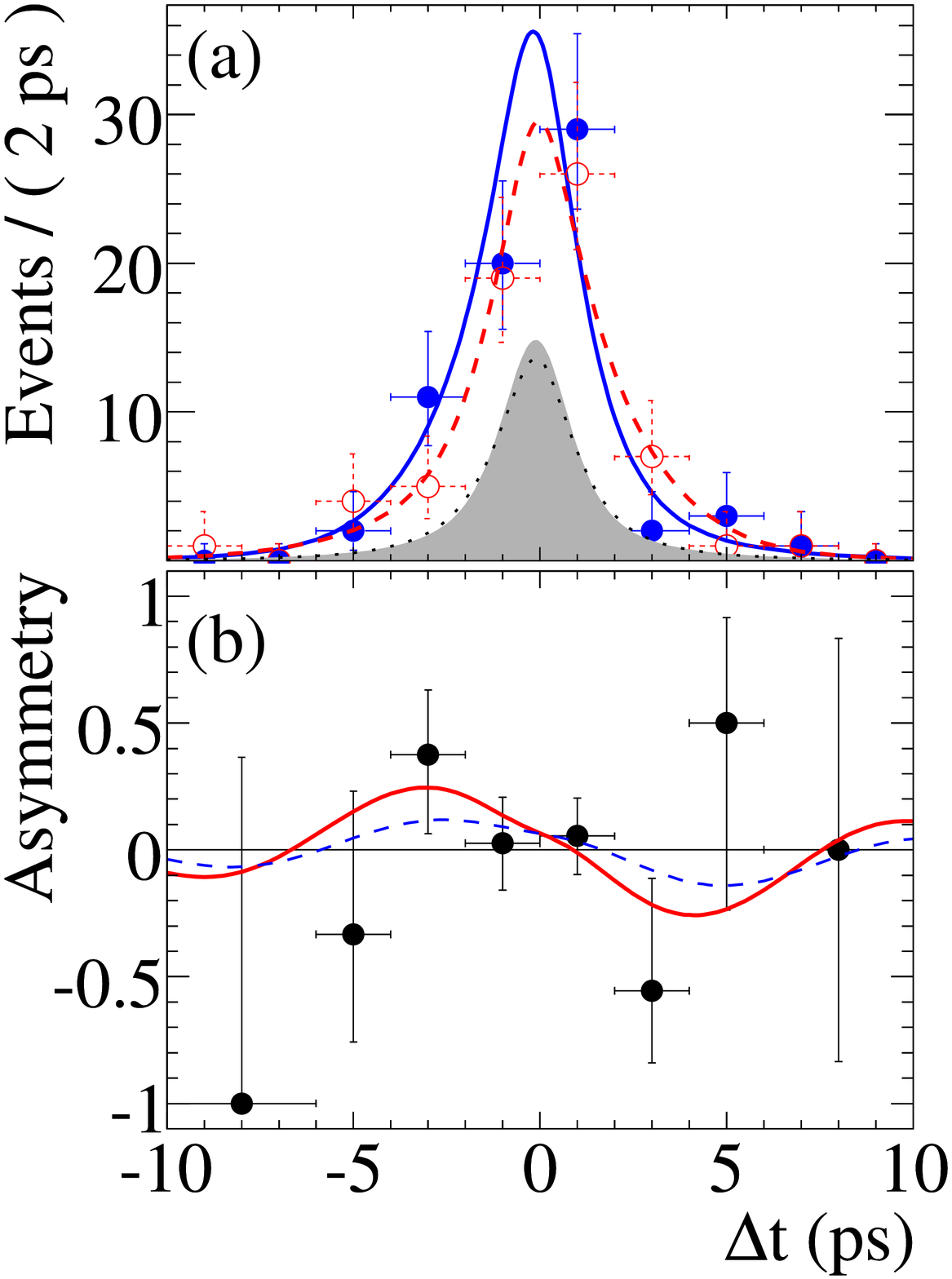}
\includegraphics[width=0.235\textwidth]{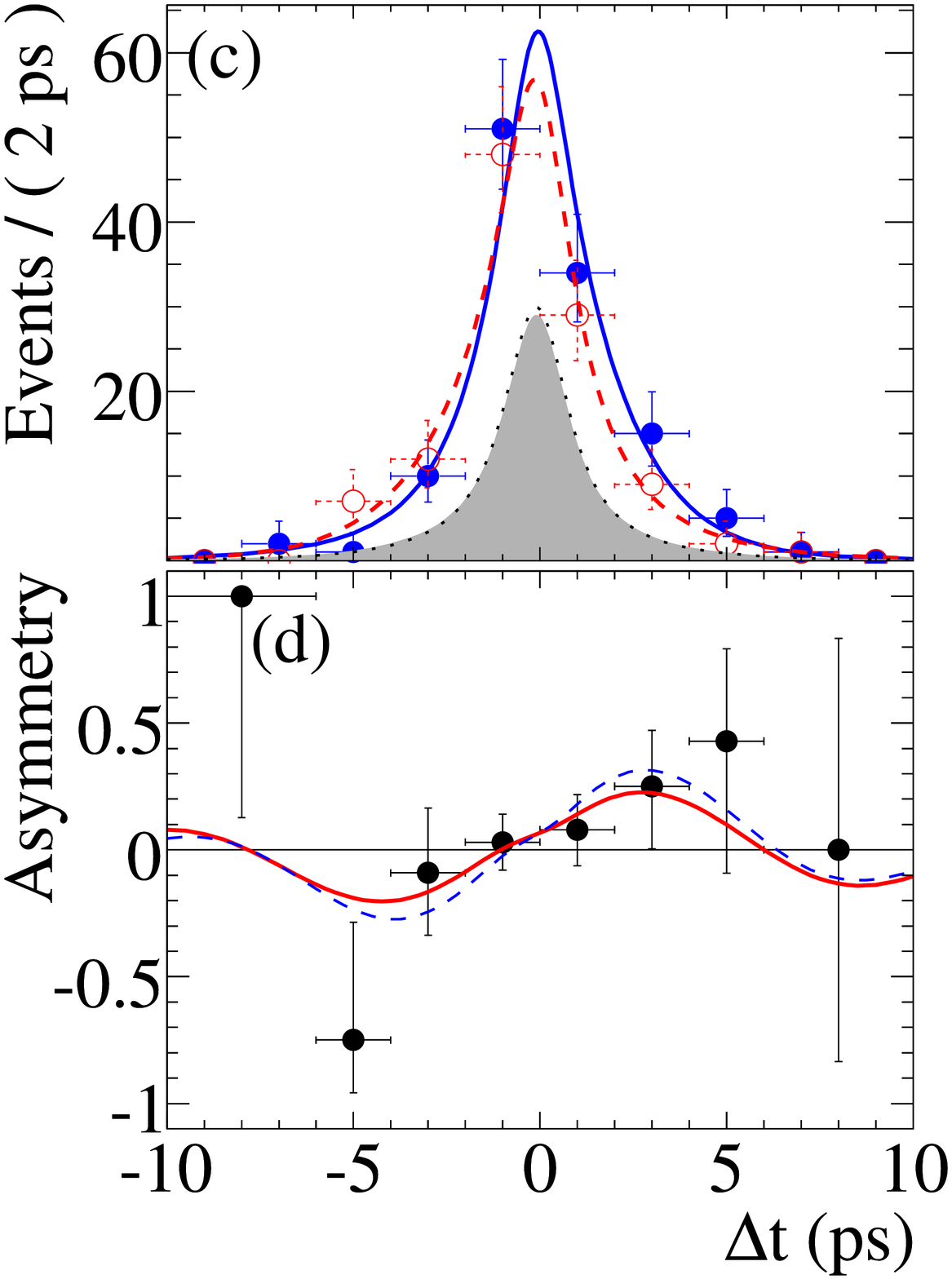}
\end{center}
\caption{The \dt distributions and asymmetries for (a,b) \CP-even and (c,d)
  \CP-odd events in the signal region ($\mES>5.27\gevcc$). In (a) and (c), the
  solid points with
  error bars and solid curve (open circles with error bars and dashed
  curve) are \Bz-tagged (\Bzb-tagged)
  data points and \dt projection curves. Shaded areas (\Bz-tagged) and
  the dotted lines (\Bzb-tagged) are background
  distributions. In (b) and (d), the solid curve represents the combined fit
  result, and the dashed curve represents the result of the fits to \CP-even
  and \CP-odd modes separately.}
\label{fig:dt-datasignal}
\end{figure}

We obtain $\calS=
-0.56 \pm 0.23 \pm 0.05$ and $\calC= -0.23\pm 0.16\pm 0.04$, where the
first errors are statistical and the second are systematic. 
The statistical correlation between \calS and \calC is $\rho=
-2.4\%$. 
The \dt distribution projections and the asymmetry
($A= [N_{\Bz\mathrm{tag}}(\dt)-N_{\Bzb\mathrm{tag}}(\dt)]/
 [N_{\Bz\mathrm{tag}}(\dt)+N_{\Bzb\mathrm{tag}}(\dt)]$) for the
events in the signal region are shown
in Fig.~\ref{fig:dt-datasignal}.
We check the consistency between  \CP-even and \CP-odd modes
by fitting them separately and find 
(statistical errors only)
$\calS_\mathrm{even}= -0.17 \pm 0.37$,
$\calS_\mathrm{odd}= -0.82 \pm 0.28$, and
$\calC_\mathrm{even}= -0.21 \pm 0.25$,
$\calC_\mathrm{odd}= -0.21 \pm 0.21$. The difference between
$\calS_\mathrm{even}$ and $\calS_\mathrm{odd}$ is $0.65\pm 0.46$, less than
1.5 standard deviation from the expected value, zero.
We also find that the differences 
 between $\hz\ra\gamma\gamma$ and $\hz\ra\pi\pi\pi$ modes are
 less than 0.1 in \calC and \calS.

The SM corrections due to the sub-leading-order diagrams are
different for $D_{\CP+}$ and $D_{\CP-}$~\cite{Fleischer}. Therefore,
we also perform a fit allowing different \CP asymmetries for $D_{\CP+}$
and $D_{\CP-}$. We obtain 
$\calS_{+} = -0.65 \pm  0.26 \pm 0.06$,
$\calC_{+} = -0.33 \pm  0.19 \pm 0.04$, 
$\rho_+=4.5\%$, and
$\calS_{-} = -0.46 \pm  0.45 \pm 0.13$, 
$\calC_{-} = -0.03 \pm 0.28 \pm 0.07$, 
$\rho_-= -14\%$.

The dominant systematic uncertainties are from the peaking background and the
\mES peak shape uncertainties (0.04 in \calS and 0.03 in \calC).
For the former,
we vary the amount of the peaking background according to its estimated
uncertainty, and vary the \CP asymmetry of the charmless component between
$\pm\sin2\beta$ of the world-average value. 
We study the latter
effect using an alternative line shape~\cite{crystal_ball} taking into
account a possible non-Gaussian tail in the \mES distribution.
Other systematic uncertainties typically do not exceed 0.01 in \calS or \calC,
and come from the following sources: the assumed parameterization of the \dt
resolution function; the uncertainties of the peaking background; \mES width
and the combinatorial background threshold function; \Bz lifetime and mixing
frequency; the beam-spot position;
and the interference between the CKM-suppressed $\bbar\ra \ubar c
\dbar $ and CKM-favored $b \ra c \ubar d$ amplitudes in some \Btag final
states, which gives deviations from the standard time evolution
function Eq.~(\ref{eq:dt})~\cite{tagsideint}.
Uncertainties due to the vertex tracker length scale and alignment
are negligible. Summing over all systematic uncertainties in
quadrature, we obtain $0.05$ for \calS and $0.04$ for \calC.

In conclusion, we have measured the time-dependent \CP asymmetry
parameters  $\calS= -0.56 \pm 0.23 \pm 0.05$ and
$\calC= -0.23 \pm 0.16\pm 0.04$ from a sample of $340\pm 32$ $\Bz\ra
D^{(*)}_{\CP}\hz$ signal events. 
The result is $2.3$ standard deviations from the \CP-conserving 
hypothesis $\calS=\calC=0$.
The parameters \calS and \calC are consistent with the SM expectation, i.e.,
the world average $-\sin2\beta= -0.725\pm0.037$~\cite{pdg} and zero,
respectively. 

We are grateful for the excellent luminosity and machine conditions
provided by our \pep2\ colleagues, 
and for the substantial dedicated effort from
the computing organizations that support \babar.
The collaborating institutions wish to thank 
SLAC for its support and kind hospitality. 
This work is supported by
DOE
and NSF (USA),
NSERC (Canada),
IHEP (China),
CEA and
CNRS-IN2P3
(France),
BMBF and DFG
(Germany),
INFN (Italy),
FOM (The Netherlands),
NFR (Norway),
MIST (Russia),
MEC (Spain), and
PPARC (United Kingdom). 
Individuals have received support from the
Marie Curie EIF (European Union) and
the A.~P.~Sloan Foundation.


\begin{thebibliography}{99}
\bibitem{RVprocess}
The $b\ra c \ubar d$ process could be mediated by a supersymmetric
$\tilde{s}_{\!R}$ in an $\not \!\!R_p$ tree process
$b \ra \ubar \tilde{s}_{\!R}$, $\tilde{s}_{\!R} \ra \c\d$.
\relax
\bibitem{conjugate}
Unless explicitly stated, charge conjugate reactions are implicitly
included throughout the paper.
\relax
\bibitem{d0tocp}
All neutral $D$ mesons in this Letter decay to \CP
eigenstates. Therefore the notation $D^{(*)0}$ implies
$D^{(*)0}_{\CP}$.
\relax
\bibitem{nominal}
All nominal masses are from~\cite{pdg}.
\relax
\end{thebibliography}
\end{document}